\documentclass[print,twocolumn,aps]{revtex4}
\usepackage{amsfonts}
\usepackage{amsmath}
\usepackage{amssymb}
\usepackage{graphicx}

\begin{document}

\title{High-power pulse trains excited by modulated continuous waves}
\author{Yan Wang$^{1}$, Lijun Song$^{2}$, Lu Li$^{1}$}
\email{llz@sxu.edu.cn}
\author{Boris A. Malomed$^{3}$}
\affiliation{$^{1}$Institute of Theoretical Physics, Shanxi University, Taiyuan 030006,
China}
\affiliation{$^{2}$College of Physics and Electronics Engineering, Shanxi University,
Taiyuan 030006, China}
\affiliation{$^{3}$Department of Physical Electronics, School of Electrical Engineering,
Faculty of Engineering, Tel Aviv University, Tel Aviv 69978, Israel}

\begin{abstract}
Pulse trains growing from modulated continuous waves (CWs) are considered,
using solutions of the Hirota equation for solitons on a finite background.
The results demonstrate that pulses extracted from the maximally compressed
trains can propagate preserving their shape and forming robust arrays. The
dynamics of double high-power pulse trains produced by modulated CWs in a
model of optical fibers, including the Raman effect and other higher-order
terms, is considered in detail too. It is demonstrated that the double
trains propagate in a robust form, with frequencies shifted by the Raman
effect.
\end{abstract}

\maketitle

\section{Introduction}

Recently, solitons on a finite background, generated by the fundamental
nonlinear Schr\"{o}dinger (NLS) equation with self-focusing, which have been
known for a long time, have drawn renewed interest as prototypes of rogue
waves in the ocean \cite%
{Kharif_Book,Kharif,Slunyaev,Zakharov,Ruban,Andonowati}, optical fibers \cite%
{Solli1,Akhmediev3,NatPhys,Kibler1,Hammani,Erkintalo}, plasmas \cite{Kwok}
and other nonlinear media. NLS solutions for solitons on a finite background
can be identified as the Kuznetsov-Ma soliton (KMS) \cite{Kuznetsov,Ma},
which is localized in the transverse direction, the Akhmediev breather (AB)
\cite{Akhmediev8}, localized in the longitudinal direction, and the
Peregrine solution (PS) \cite{Peregrine}, which is a limit case of the KMS
and AB, being localized in both the transverse and longitudinal directions.
The spatiotemporal localization predicted by the PS solution, and dynamical
manifestations of the KMS have been demonstrated in optical fibers \cite%
{NatPhys,Kibler1}. Splitting of the PS produced by deformed initial
conditions, and the related modulational instability have been observed too
\cite{Hammani,Erkintalo}. The dynamics accounted for by the AB was first
observed in Ref. \cite{Dudley1} and further analyzed in Ref. \cite{NatPhys}.
Collisions between ABs \cite{Frisquet1}, their dynamics in optical fibers
with a longitudinally tailored dispersion \cite{OL39_4490}, and shaping of
an optical frequency comb into a generator of rogue waves \cite{Frisquet2}
have been demonstrated too. In these experiments, the ABs and KMSs were
excited by weakly and strongly modulated continuous waves (CWs),
respectively.

Generally, NLS solutions initiated by perturbations added to a finite
background feature strong temporal compression of emerging peaks \cite{Ygy}.
In optical systems, such dynamical scenarios can be used for the generation
of supercontinuum, and of high-power ultrashort pulses and pulse trains \cite%
{Dudley,Solli3,Mussot,Dudley1,Taki}.

To create high-repetition-rate and high-quality pulse trains, the delay-line
interferometer was used to eliminate the finite background around the AB and
allow doubling of the repetition rate of the pulse trains \cite{Finot}. A
possibility of spontaneous transformation of ABs into solitons and
ultrashort pulses generated by the modulation instability have been
considered too \cite{PRA85_033808,APB9_2013}. Extraction of high-power
optical pulses from the PS by means of the spectral-filtering method was
proposed in Refs. \cite{Ygy1,Ygy2,Ygy3}. However, these results are only
suitable for picosecond pulses with the width $\gtrsim 5$ ps \cite%
{Kodama,Agrawal}. In the present paper, using soliton solutions on the
finite background for the Hirota equation, we construct high-power strongly
compressed pulse trains induced by modulated CWs in the subpicosecond
regime, an appropriate model for which must include higher-order terms,
added to the usual NLS equation.

The paper is organized as follows. In Sec. II, the Hirota equation, which
models the pulse propagation in the nonlinear fiber, is presented, and
solutions for solitons on the finite background are recapitulated for this
equation. In the same section, using AB and KMS solutions for the Hirota
equation, high-power compressed pulse trains induced by modulated CWs are
demonstrated in the model with the higher-order terms, making use of
juxtaposition of two solutions with a half-period temporal shift and
opposite signs. The influence of the Raman effect on the high-power pulse
trains is considered in Sec. III. Conclusions are summarized in Sec. IV.

\section{High-power pulse trains produced by the Hirota equation}

As said above, the standard NLS equation is valid as the model for
picosecond pulses with the width $\gtrsim 5$ps, while in the subpicosecond
or femtosecond regime higher-order terms should be included, to account for
the third-order dispersion (TOD), self-steepening, and the Raman effect \cite%
{Kodama,Agrawal}. This accordingly modified NLS equation is%
\begin{gather}
\text{ \ \ \ }\frac{\partial A}{\partial z}+i\frac{\beta _{2}}{2}\frac{%
\partial ^{2}A}{\partial T^{2}}-\frac{\beta _{3}}{6}\frac{\partial ^{3}A}{%
\partial T^{3}}  \notag \\
=i\gamma \left[ |A|^{2}A+\frac{i}{\omega _{0}}\frac{\partial |A|^{2}A}{%
\partial T}-(i\alpha _{\nu }+T_{R})A\frac{\partial |A|^{2}}{\partial T}%
\right] .  \label{model}
\end{gather}%
Here $A=A(z,T)$ is the slowly varying envelope of the electric field, $%
T=t-z/v_{g}\equiv t-\beta _{1}z$, where $t$ and $z$ are the temporal
variable and propagation distance, and $v_{g}$ is the group velocity.
Coefficients $\beta _{2}$, $\beta _{3}$, $\gamma $, $\alpha _{\nu }$ and $%
T_{R}$ account for the second-order group-velocity dispersion (GVD), TOD,
strength of the Kerr nonlinearity, nonlinear dispersion, and the Raman
time-delay constant, respectively. Equation (\ref{model}) does not include
the fiber loss, assuming, as usual, that it may by compensated by gain \cite%
{Kodama,Agrawal}. At the end of Sec. III, we produce results of the analysis
of the model which explicitly includes the loss, the conclusion being that
the difference is relatively small except for a decreasing of peak power.

By means of rescaling, $A(z,T)\equiv \sqrt{P_{0}}q(\xi ,\tau )$, $\tau
\equiv T/T_{0}$, and $\xi =z/L_{D}$, with the temporal scale $%
T_{0}=[\left\vert \beta _{2}\right\vert /(\gamma P_{0})]^{1/2}$ and
dispersion length $L_{D}=(\gamma P_{0})^{-1}$, Eq. (\ref{model}) is
transformed into
\begin{gather}
\frac{\partial q}{\partial \xi }=i\left( \frac{s}{2}\frac{\partial ^{2}q}{%
\partial \tau ^{2}}+\left\vert q\right\vert ^{2}q\right) +\alpha _{3}\frac{%
\partial ^{3}q}{\partial \tau ^{3}}  \notag \\
+\alpha _{4}\frac{\partial \left\vert q\right\vert ^{2}q}{\partial \tau }%
+\left( \alpha _{5}-i\tau _{R}\right) q\frac{\partial \left\vert
q\right\vert ^{2}}{\partial \tau },  \label{He}
\end{gather}%
where $s\equiv -\beta _{2}/\left\vert \beta _{2}\right\vert $ is $+1$ and $%
-1 $ for the anomalous and normal GVD, respectively, and the other
coefficients are
\begin{equation}
\alpha _{3}=\frac{\beta _{3}}{6\left\vert \beta _{2}\right\vert T_{0}}%
,\alpha _{4}=-\frac{1}{\omega _{0}T_{0}},\alpha _{5}=\frac{\alpha _{\nu }}{%
T_{0}},\tau _{R}=\frac{T_{R}}{T_{0}},  \label{alpha}
\end{equation}%
which, as well as their unscaled counterparts, are related to the TOD,
self-steepening, nonlinear dispersion, and the Raman effect, respectively.
In its general form, Eq. (\ref{He}) is not integrable. However, if the Raman
effect is omitted, i.e., $\tau _{R}=0$, and additional conditions
\begin{equation}
\alpha _{4}=6\alpha _{3},~\alpha _{4}+\alpha _{5}=0,  \label{int}
\end{equation}%
are satisfied, Eq. (\ref{He}) becomes the integrable Hirota equation \cite%
{Hirota}. Here we only consider the case of the anomalous GVD, i.e. $s=1$.
In this case, the solution of the Hirota equation for a soliton placed on
top of a finite background can be compactly written as \cite%
{Ygy,Lishuqing,Ankiewicz}%
\begin{equation}
q(\xi ,\tau )=\left[ 1+\frac{2\left( 1-2a\right) \cosh (b\xi )+ib\sinh (b\xi
)}{\sqrt{2a}\cos [\omega (\tau +c\xi )]-\cosh (b\xi )}\right] e^{i\xi },
\label{AB}
\end{equation}%
\begin{equation}
\omega =2\sqrt{1-2a},~b=\sqrt{8a(1-2a)},~c=2(1+4a)\alpha _{3},  \label{abc}
\end{equation}%
where $a$ is a real constant parametrizing this class of the solutions \cite%
{Kibler1}. When $\alpha _{3}=0$, solution (\ref{AB}) goes over into its NLS
counterpart \cite{Kibler1}.

For $0<a<1/2$, constants $\omega $, $b$ and $c$ in Eq. (\ref{AB}) are real.
In this case, the solution describes the AB, and the maximally compressed
pulse train is attained at point $\xi =0$, in the form of
\begin{equation}
q_{\max }(\tau )=\frac{1-4a+\sqrt{2a}\cos (\omega \tau )}{\sqrt{2a}\cos
(\omega \tau )-1},  \label{peak}
\end{equation}%
with period $2\pi /\omega $ and background power
\begin{equation}
P_{\text{bg}}=(2\sqrt{2a}-1)^{2},  \label{bg}
\end{equation}%
which corresponds to $\cos (\omega \tau )=-1$. The analysis reported in Ref.
\cite{OL39_4490} has demonstrated that, in the range of $0<a<0.2$, the
evolution of the maximally compressed pulse train can be frozen with the
help of a specifically tailoring the GVD profile, which leads to formation
of a quasi-stable train of fundamental solitons, while at $a>0.2$ a
delay-line interferometer can be used to create a high-power pulse train
with the double repetition rate and zero background, as shown in Ref. \cite%
{Finot}.

A natural question is whether such a double-repetition-rate high-power train
can stably propagate in the framework of the fiber model. To address this
issue, it is necessary to choose suitable initial conditions, which can
excite a maximally compressed pulse array similar to the AB profile (\ref%
{peak}). It has been shown that such a maximally compressed pulse train can
be excited by a modulated CW \cite{PLA375_2029,Hammani}, corresponding to
the initial condition%
\begin{equation}
q(0,\tau )=\sqrt{1+\delta \cos \left( \omega \tau \right) },
\label{AB_initial}
\end{equation}%
where $\delta $ is a small modulation amplitude, and $\omega $ is the
modulation frequency, taken as per Eq. (\ref{abc}).

\begin{figure}[tbp]
\centering\vspace{0.0cm} \includegraphics[width=8.5cm]{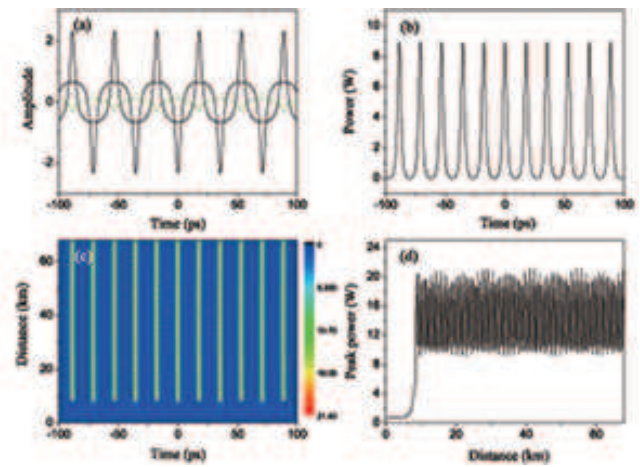} \vspace{%
-0.0cm}
\caption{(Color online) (a) The maximally compressed pulse train at $%
z=8.4206 $ km, juxtaposed with its counterpart with the opposite sign,
delayed by a half a period. Here the blue dotted and green dashed-dotted
curves depict, respectively, the real and imaginary parts of the numerical
result, while the black solid curve shows the exact solution, which actually
overlaps with the blue dotted curve. (b) The intensity distribution of the
resulting double pulse train, produced as the superposition of the two
original trains. (c) The evolution of the double train with zero background.
(d) The evolution of the peak power of the double train with zero
background. These results correspond to $a=0.4$ in Eq. (\protect\ref{abc})
and $\protect\delta =0.01$ in Eq. (\protect\ref{AB_initial}).}
\label{fig1}
\end{figure}

We numerically simulated the integrable Hirota equation with the initial
modulated CW state (\ref{AB_initial}), where the system parameters are
adjusted to the SMF-28 fiber with $\beta _{2}$ $=-21.4$ ps$^{2}$km$^{-1}$, $%
\beta _{3}=0.12$ ps$^{3}$km$^{-1}$, and $\gamma =1.2$ W$^{-1}$km$^{-1}$
around $1550$ nm \cite{Hammani}, and the initial power $P_{0}=0.7$ W. The
simulations are necessary, as generic solutions of equations cannot be
obtained in an explicit analytical form. In fact, high-power pulse trains
growing from the modulated CW do not necessarily reduce to exact breather
solutions of the Hirota equation, such as KMS \cite{Finot}. In the present
case, the normalized parameters are $T_{0}=$ $5.0474$ ps, $L_{D}=1.1905$ km,
and $\alpha _{3}=1.8516\times 10^{-4}$, respectively. The simulation
demonstrates that the maximally compressed pulse train is attained at $%
z=8.4206$ km, as shown in Fig. \ref{fig1}(a). At this position, we superpose
the train and its counterpart with the opposite sign, delayed by half a
period, to produce a double pulse train, which is displayed in Fig. \ref%
{fig1}(b). The procedure can be experimentally realized by using a delay
line and a piezo-electrical device to impose the necessary delay of the
pulse train and the $\pi $ phase-shift, as shown in Ref. \cite{Finot}. Then,
we let the resulting pulse train propagate in the fiber, see the
corresponding stable regime in Fig. \ref{fig1}(c). Unlike its exact
counterpart, the numerical solution at the point of the strongest
compression includes an imaginary part [shown by the green dashed-dotted
curves in Fig. \ref{fig1}(a)], whose maximum value is $0.0985\approx 1\%$ of
the amplitude in the real part, which is a small perturbation that was
included to test the stability of the propagation of the pulse train with
the zero background. Because the soliton number, $N\equiv \lbrack \gamma
P(\Delta \tau /1.763)^{2}/\left\vert \beta _{2}\right\vert )]^{1/2}$, takes
value $1.3617$ for the present parameter set, which exceeds $1$, individual
pulses in the resulting train, with peak power $P$ and full-width at
half-maximum (FWHM) $\Delta \tau =3.3878$ps, exhibit oscillations, as shown
in Fig. \ref{fig1}(d). We stress that the oscillatory regime is a
dynamically stable one, as is clearly shown by the simulations.

Next, we turn to the case of $a>1/2$, which corresponds to the high
background power, as per Eq. (\ref{bg}). In this case, Eq. (\ref{abc})
yields imaginary $\omega $ and $b$, converting the hyperbolic and
trigonometric functions in Eq. (\ref{AB}) into trigonometric and hyperbolic
ones, respectively, thus swapping localization and periodicity features of
the solution. It now represents a bright oscillating pulse propagating on
top of a finite background, which is generally classified as KMS.
Accordingly, the maximally compressed pulse is attained at $\xi =0$%
\begin{equation}
\ q_{\max }(\tau )=\dfrac{1-4a+\sqrt{2a}\cosh (\widetilde{\omega }\tau )}{%
\sqrt{2a}\cosh (\widetilde{\omega }\tau )-1},  \label{Ma-pulse}
\end{equation}%
with $\widetilde{\omega }=2(2a-1)^{1/2}$. This pulse has a single peak at $%
\tau =0$ [$q_{\max }(0)=-(1+2\sqrt{2a})$], and background value $q=1$ as $%
|\tau |\rightarrow \infty $, which makes the solution completely different
from the AB.

In principle, the maximally compressed pulse can be created by a small
localized sech-type perturbation added to a CW background \cite{Ygy}. On the
other hand, it was demonstrated in Ref. \cite{Kibler1} that such a pulse can
be excited by a strongly modulated CW in the form of%
\begin{equation}
q(0,\tau )=1+A\left[ 1+\cos \left( \Omega \tau \right) \right]
\label{Ma-initial}
\end{equation}%
with $\Omega =\widetilde{\omega }\cos ^{-1}[(\sqrt{4a-2\sqrt{2a}+1}-1)/(%
\sqrt{2a}-1)-1]/\cosh ^{-1}\{1[(4a-2)/(\sqrt{4a-2\sqrt{2a}+1}-1)-1]/\sqrt{2a}%
\}$ and $A=\sqrt{2a}-1$, which has the peak power and FWHM identical to
those of the intensity profile, $|q_{\min }(\tau )|^{2}=\{1+[2(2a-1)]/[\sqrt{%
2a}\cosh (\widetilde{\omega }(\tau +c\pi /\widetilde{b}))+1]\}^{2}$, at the
point of the minimal intensity, $\xi =\pi /\widetilde{b}$, where $\widetilde{%
b}=[8a(2a-1)]^{1/2}$.

\begin{figure}[tbp]
\centering\vspace{0.0cm} \includegraphics[width=8.5cm]{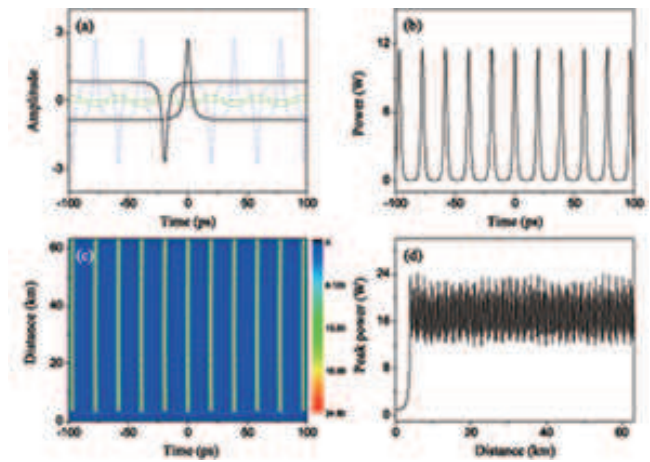} \vspace{%
-0.0cm}
\caption{(Color online) (a) The maximally compressed pulse train at $%
z=3.7080 $ km and its counterpart with the opposite sign and temporal delay $%
\protect\pi /\Omega $. (b) The intensity distribution of the train produced
by the juxtaposition of both original trains. (c) The evolution of the
resulting pulse train with zero background. (d) The evolution of the peak
power of the resulting pulse train with zero background. Here $a=0.6$. The
notation adopted in the panels is the same as in Fig. \protect\ref{fig1}. }
\label{fig2}
\end{figure}

Starting from input (\ref{Ma-initial}), we have simulated, similar to what
is shown above in Fig. \ref{fig1}, the evolution produced by the
juxtaposition of the maximally compressed pulse trains and its counterpart
with the opposite sign and half-a-period delay. The results, summarized in
Fig. \ref{fig2}, show that, for our choice of the parameters, the maximally
compressed pulse train is attained at $z=3.7080$ km [Fig. \ref{fig2}(a)],
with individual pulses in the train being almost identical to the profile of
the maximally compressed pulse given by Eq. (\ref{Ma-pulse}). Thus,
combining the pulse train with its counterpart with the opposite sign and
time delay $\pi /\Omega $, one can construct a double pulse train with zero
background [see Fig. \ref{fig2}(b)]. It propagates stably, exhibiting
oscillatory behavior of the pulses, as the respective soliton number is $%
N=1.2761$, with FWHM $\Delta \tau =2.7927$ ps, which again exceeds $1$, see
Figs. \ref{fig2}(c) and (d). Similarly, unlike the exact solution, the
numerical result includes an imaginary part [see the green dash dotted
curves in Fig. \ref{fig2}(a)], with maximum value $0.0583\approx 0.76\%$ of
the amplitude of the real part, which does not affect the stable propagation
of the pulse train with zero background.

Finally, it should be pointed out that, in the case of $a\rightarrow 1/2$,
solution (\ref{AB}) of the Hirota equation reduces to the PS in the form of
\begin{equation*}
q(\xi ,\tau )=\left[ 1-\frac{4\left( 1+2i\xi \right) }{1+4\xi ^{2}+4(\tau
+6\alpha _{3}\xi )^{2}}\right] e^{i\xi },
\end{equation*}%
which is a superposition of the CW solution and a rational fraction
function, forming a maximally compressed pulse at $\xi =0$ \cite%
{Peregrine,Ygy1}. This implies that the Peregrine rogue wave, i.e., the
maximally compressed pulse, can be excited by a small localized
(single-peak) perturbation placed on top of the CW background \cite{Ygy}.
Thus, the high-power pulse extracted from the Peregrine rouge wave can be
realized by means of the spectral-filtering method, displaying a
breather-like behavior \cite{Ygy2,Ygy3}.

\section{The influence of the Raman effect on high-power pulse trains}

In the previous Section, we have considered the generation and propagation
of the pulse trains growing from the CWs as solutions of the Hirota
equation, i.e., Eq. (\ref{He}) under the integrability relations (\ref{int}%
), while the Raman effect was not be included ($\tau _{R}=0$). This model
does not precisely apply to real optical fibers, but the exact solutions
provided by it help one to understand the pulse-train dynamics in more
general models.

In this section, we report results of simulations of Eq. (\ref{He}) with
realistic values of parameters (\ref{alpha}), including a nonzero Raman
coefficient, and without imposing integrability conditions (\ref{int}): $%
\alpha _{3}=1.8516\times 10^{-4}$, $\alpha _{4}=-1.6292\times 10^{-4}$, $%
\alpha _{5}$ is determined by the delayed-nonlinear-response parameter $%
\alpha _{\nu }$, and $\tau _{R}=5.9437\times 10^{-4}$ for the usual value of
the Raman time constant, $T_{R}=3$ fs. Actually, $\alpha _{\nu }$ is a small
quantity which has little effect on simulation results, therefore we set $%
\alpha _{\nu }=0$. Thus, we focus on the influence of the Raman effect on
the dynamics of the high-power pulse trains growing from the modulated CWs
introduced as per Eqs. (\ref{AB_initial}) and (\ref{Ma-initial}).

In the analysis presented above the modulation amplitude $A$ and frequencies
$\omega $ and $\Omega $ in Eqs. (\ref{AB_initial}) and (\ref{Ma-initial})
were taken as functions of parameter $a$, making it possible to compare the
results with the exact solutions produced by the Hirota equation, as shown
in Fig. \ref{fig1}(a) and Fig. \ref{fig2}(a). However, in the realistic
fiber-optic model, the modulation amplitudes and frequencies can take
arbitrary values. Here, we use inputs given by Eqs. (\ref{AB_initial}) and (%
\ref{Ma-initial}) with different values of the parameters, aiming to excite
the maximally compressed pulse trains, and perform subsequent simulations of
the corresponding double pulse trains in the framework of Eq. (\ref{He}).

\begin{figure}[tbp]
\centering\vspace{0.0cm} \includegraphics[width=8.5cm]{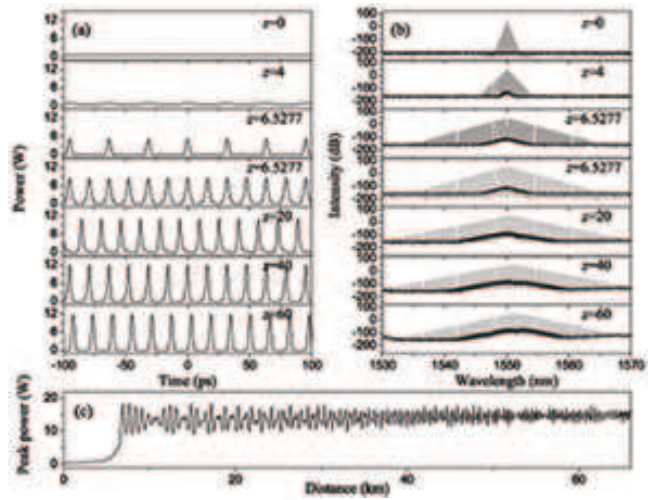} \vspace{%
-0.0cm}
\caption{(Color online) (a) Power profiles at different propagation
distances for the pulse train generated by the modulated CW given by Eq. (%
\protect\ref{AB_initial}). (b) The corresponding power spectra. (c) The
evolution of the peak power. Here the parameters are $\protect\delta =0.03$,
$\protect\omega =1$, $T_{R}=3$ fs, and $\protect\alpha _{\protect\nu }=0$.}
\label{fig3}
\end{figure}

\begin{figure}[tbp]
\centering\vspace{0.0cm} \includegraphics[width=8.5cm]{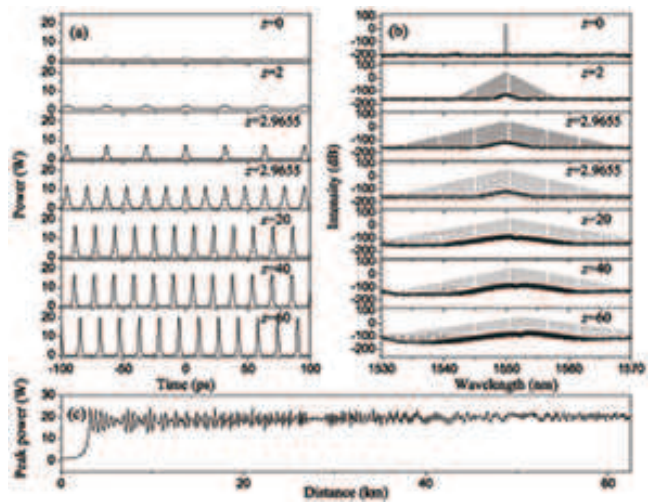} \vspace{%
-0.0cm}
\caption{(Color online) The same as in Fig. \protect\ref{fig3}, but for the
pulse train generated by the input in the form of Eq. (\protect\ref%
{Ma-initial}). The parameters are $A=0.13$, $\Omega =1$, $T_{R}=3$ fs, and $%
\protect\alpha _{\protect\nu }=0.$}
\label{fig4}
\end{figure}

Fig. \ref{fig3} shows the evolution of the train initiated by input (\ref%
{AB_initial}) with $\delta =0.03$ and $\omega =1$. In Fig. \ref{fig3}(a) one
can see that the modulated CW initially undergoes strong temporal
compression and increase of the peak power, attaining the maximally
compressed state of the train at $z=6.5277$ km. At that stage, the double
pulse train is constructed, using the delay-line interferometer (the method
is outlined in the previous Section, see also Ref. \cite{Finot}), which
robustly propagates along the fiber. Fig. \ref{fig3}(b) displays the
concomitant evolution of the pulse-train's power spectrum, showing that the
spectrum initially expands until the point of the maximum compression, at
which the spectral distribution is reduced to a half of the original
spectrum \cite{Finot}. Subsequently, the spectral width remains essentially
constant, while the frequencies feature the red shift induced, as usual, by
the Raman effect \cite{Kodama,Agrawal}. For the presently chosen parameters,
the frequency shift of the train is about $1.1967$ THz at $60$ km. In
addition, the evolution of the peak power, which exhibits an oscillatory
behavior, is displayed in Fig. \ref{fig3}(c).

Fig. \ref{fig4} displays the evolution initiated by the input given by Eq. (%
\ref{Ma-initial}) with $A=0.13$ and $\Omega =1$. This scenario exhibits
three differences in comparison with Fig. \ref{fig3}. One is that the
maximally compressed pulse train is attained at a shorter distance, i.e., $%
z=2.9655$ km, as seen in Fig. \ref{fig4}(a). The other is that the initial
spectrum is narrower, while the Raman-induced red frequency shift is larger,
as seen in Fig. \ref{fig4}(b), in which the frequency shift of the train is
about $2.3854$ THz at $60$ km. The third difference is a higher peak power,
as seen in Fig. \ref{fig4}(c).

\begin{figure}[tbp]
\centering\vspace{0.0cm} \includegraphics[width=8.5cm]{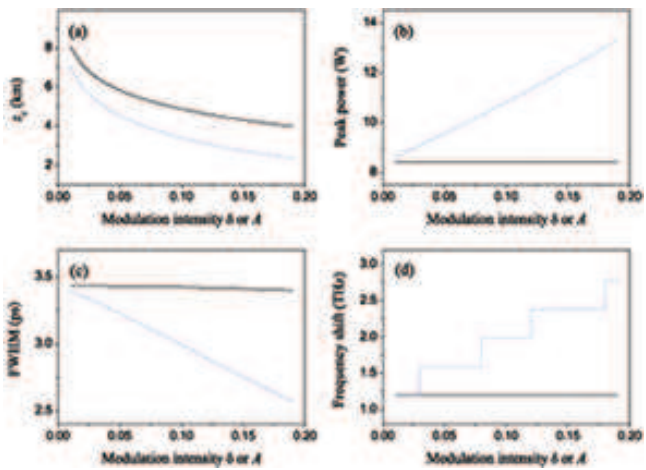} \vspace{%
-0.0cm}
\caption{(Color online) (a) The position of the maximally compressed pulse
train, (b) the peak power, (c) the FWHM\ of the central pulse in the double
pulse train, and (d) the Raman-induced frequency shift after passing $60$ km
in the fiber, versus the modulation intensity $\protect\delta $ or $A$, with
the black solid and blue dotted curves pertaining to inputs (\protect\ref%
{AB_initial}) and (\protect\ref{Ma-initial}), respectively. Here the
parameters are $\Omega =\protect\omega =1$, $T_{R}=3$ fs, and $\protect%
\alpha _{\protect\nu }=0.$}
\label{fig5}
\end{figure}

To summarize the dynamical results outlined above, in Fig. \ref{fig5} we
present the largest-compression position $z_{c}$ for the pulse trains, their
peak power, the FWHM of the central pulse in the double pulse trains
constructed with the help of the delay-line interferometer method, and the
corresponding frequency shift accumulated after the $60$ km long
propagation, as functions of parameters $\delta $ and $A$ in inputs (\ref%
{AB_initial}) and (\ref{Ma-initial}), respectively.

In Fig. \ref{fig5}(a), one can see that, for the same modulation intensity,
the position of the strongest compression corresponding to initial condition
(\ref{Ma-initial}) is always smaller than its counterpart corresponding to
input (\ref{AB_initial}). Interestingly, for the double pulse train induced
by the latter input, the peak power and the FWHM of the central pulse are
almost constant for the present parameters, which leads to the nearly
constant frequency shift at $60$ km. On the other hand, for the double pulse
train induced by input (\ref{Ma-initial}), the peak power increases, and the
FWHM decreases, with the increase of the modulation intensity, which implies
that this input can be used to generate the pulse train with the higher peak
power and narrower width. However, in the same case, the frequency shift at $%
60$ km increases with the increase of the modulation intensity in a
staircase fashion, as shown in Fig. \ref{fig5}(d). Furthermore, from Figs. %
\ref{fig5}(b) and (c) one can find that the corresponding soliton numbers
are in the range of $1.2592<N<1.3412$, each pulse in the double pulse trains
eventually reshaping into the fundamental soliton, as seen in Figs. \ref%
{fig3}(c) and \ref{fig4}(c). Comparing with Figs. \ref{fig1}(d) and \ref%
{fig2}(d), one can see that Raman effect is helpful for the formation of
solitons over a shorter propagation distance, as was shown in a different
context in Ref. \cite{Mamyshev}.

\begin{figure}[tbp]
\centering\vspace{-0.0cm} \includegraphics[width=8.5cm]{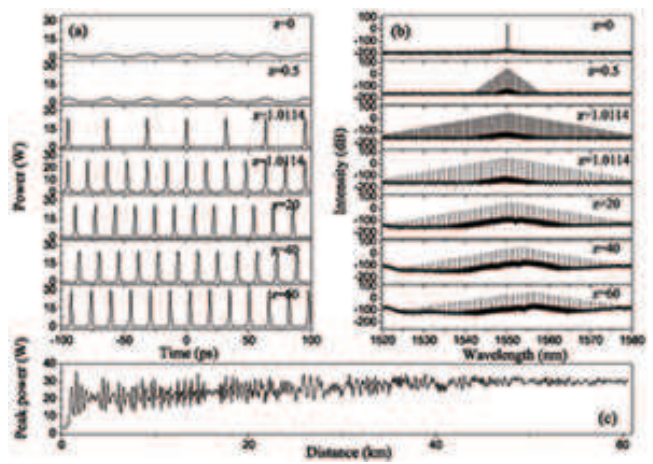} \vspace{%
-0.0cm}
\caption{(Color online) The same as in Fig. \protect\ref{fig4} except for $%
A=0.55$.}
\label{fig6}
\end{figure}

From Fig. \ref{fig5}(c) it can be seen that the pulse width in the trains is
less than $2$ ps (here the pulse width is taken equal to FWHM/$1.763$ \cite%
{Agrawal}), and it may fall in the subpicosecond range for larger values of
modulation intensity $A$. These widths are small enough to make it necessary
using the extended model based on Eq. (\ref{model}), rather than the simple
NLS equation which does not include higher-order terms. To additionally
illustrate this point, we have performed simulations of the propagation
dynamics of the train initiated by input (\ref{Ma-initial}) with $A=0.55$ in
the subpicosecond range. In this case, the maximally compressed pulse train
is attained at a shorter distance $z=1.0114$ km in comparison with Fig. \ref%
{fig4}. Subsequently, the use of the delay-line interferometer produces the
double pulse train with the FWHM of the central pulse $1.4083$ ps (the
corresponding pulse width is $\allowbreak 799$ fs) and peak power $26.3165$
W. The corresponding robust evolution of the double pulse train is shown in
Fig. \ref{fig6}(a). Comparing with the results shown in Fig. \ref{fig4}, we
conclude that the spectrum range is wider, the frequency shift is larger,
and the peak power is higher because the pulse width becomes narrower, as
seen in Figs. \ref{fig6}(b) and (c).

\begin{figure}[tbp]
\centering\vspace{-0.0cm} \includegraphics[width=8.5cm]{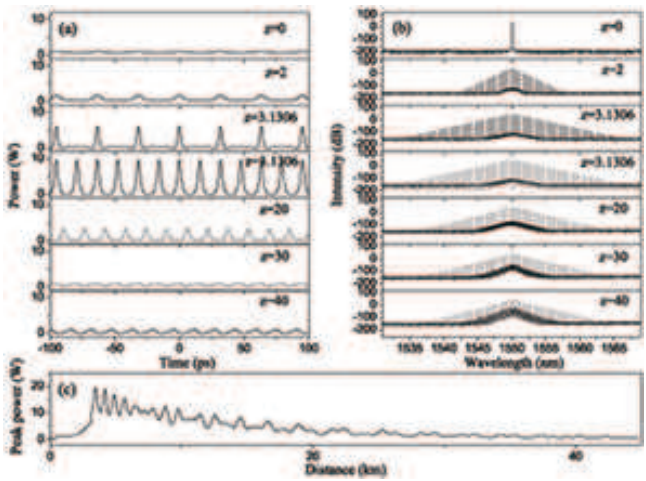} \vspace{%
-0.0cm}
\caption{(Color online) The same as in Fig. \protect\ref{fig4}, but with
fiber loss $0.19$ dB/km.}
\label{fig7}
\end{figure}

Note that the above considerations disregarded fiber loss, assuming that it
might be compensated by gain. Here, we display an example which explicitly
takes the loss into account, and the results are shown in Fig. \ref{fig7},
in which the maximally compressed pulse train is attained at a somewhat
longer distance, i.e., $z=$\ $3.1306$ km (it was $2.9655$km without fiber
loss) and the peak power of the corresponding double pulse train is
decreased to $9.6711$ W, from $11.6262$ W in the lossless fiber. Comparing
with Fig. \ref{fig4}, we see that the propagation scenario is not strongly
affected by the loss. Of course, the loss produces stronger effects over
large propagation distances ($\gtrsim 30$ km), in which case periodic
compensation provided by gain should be explicitly taken into account.

\section{Conclusions}

Using exact solutions of the Hirota equation for the soliton on the finite
background, including the AB (Akhmediev breather) and KMS (Kuznetsov-Ma
solution), the double high-power pulse trains with zero-background have been
constructed. The results demonstrate that the double trains, built as
superpositions of the maximally compressed pulse trains with opposite signs
and a half-period time delay, propagate preserving their shape. The dynamics
of the trains generated by the modulated CWs in the realistic fiber model
has been simulated too, demonstrating that they can propagate robustly, with
the red frequency shift caused by the Raman effect. Comparing different
initial conditions, we have concluded that input (\ref{Ma-initial}) is more
suitable for generating the pulse trains with a higher peak power and
narrower width, the corresponding Raman-induced frequency shift being larger
too, exhibiting a staircase dependence on the CW-modulation parameters. The
settings analyzed in this work suggest an experimental realization in
nonlinear optical fibers.

\section{Acknowledgment}

This research is supported by the National Natural Science Foundation of
China grant 61078079 and 61475198, the Shanxi Scholarship Council of China
grant 2011-010.

\end{document}